\documentstyle[aps,multicol,eqsecnum,epsfig]{revtex}

\newcommand\be{\begin{eqnarray}}
\newcommand\ee{\end{eqnarray}}
\newcommand\ba{\begin{array}}
\newcommand\ea{\end{array}}
\def\r{\rangle}
\def\l{\langle}

\def\cH{{\cal H}}

\def\openone{{\it I}}

\begin{document}
\title{Improving performance of probabilistic programmable quantum processors}
\author{Mark Hillery${}^{1}$, M\'ario Ziman${}^{2}$, and Vladim\'\i r Bu\v zek${}^{2,3}$}
\address{
${}^{1}$Department of Physics, Hunter College of CUNY,
 695, Park Avenue, New York, NY 10021, U.S.A.\\
${}^{2}$ Research Center for Quantum Information,
Slovak Academy of Sciences,
D\'ubravsk\'a cesta 9, 845 11 Bratislava, Slovakia \\
${}^{3}$Faculty of Informatics, Masaryk University, Botanick\'a 68a,
602 00 Brno, Czech Republic
}
\maketitle
\begin{abstract}
We present a systematic analysis how one can improve performance of probabilistic programmable quantum
processors. We generalize
 a simple Vidal-Masanes-Cirac processor that realizes U(1) rotations on a qubit
with the phase of the rotation encoded in a state of the program register. We show how the probability
of success of the probabilistic processor can be enhanced by using the processor in loops. In addition
we show that the same strategy can be utilized for a probabilistic implementation of non-unitary
transformations on qubits.
In addtion, we show that an arbitrary
SU(2) transformations of qubits can be encoded in program state of a universal programmable
probabilistic quantum processor. The probability of success of this processor can be enhanced by
a systematic correction of errors via conditional loops.
Finally, we show that all our results can be generalized also for qudits. In particular,
we show how to implement SU (N) rotations of qudits via programmable quantum processor
and how the performance of the processor can be enhanced when it is used in loops.
\end{abstract}


\begin{multicols}{2}
\section{Introduction}
\label{sec1}
Classical computers are programmable, that is, the task they perform
is determined by a set of instructions that is sent into the machine
along with the data to be processed.  This is a very desirable feature;
we do not have to build a different circuit every time we want to
perform a different procedure.  It would be useful to be able to develop
quantum processors with the same property.

The development of programmable quantum circuits is an area that has
attracted attention only  recently.  The basic model for these circuits
consists of two parts, a data register and a program register.  There
are two inputs, a data state, which is sent into the data register, and
on which an operation is to be performed, and a program state, which is
sent into the program register,  that specifies the operation.
The first result was due to
Nielsen and Chuang, who showed that a deterministic {\it universal}
quantum processor does not exist \cite{Nielsen1997}.  The problem is
that a new dimension must be added to the program space for each
unitary operator that one wants to be able to perform on the data.
A similar situation holds if one studies quantum circuits that
implement completely-positive, trace-preserving maps rather than
just unitary operators \cite{Hillery2001,Hillery2002a}. Some families of maps can
be implemented with a finite program space, for example, the
phase damping channel, but others, such as the amplitude damping
channel, require and infinite program space.
If one drops the requirement that the processor be deterministic,
then universal processors become possible
\cite{Nielsen1997,Preskill1998,Vidal2002,Hillery2002b}.
These processors are probabilistic: they sometimes fail, but we
know when this happens.

A number of examples of programmable quantum circuits have been
proposed. One is a quantum ``multimeter'' that performs unambiguous
state discrimination on a set of two states, the set being
specified by the program \cite{Dusek2002}.  There are also
devices that evaluate the expectation value of an arbitrary operator,
the data representing the state in which the expectation value is
to be evaluated and the program state specifying the operator
\cite{Ekert2002,Paz2003}.

In a probabilistic processor, one measures  the output program state.
If the proper result is obtained, the desired operation has been
performed on the data state, and if not, then the output of the data
register is discarded. In this kind of a scenario, one wants the
probability of successfully performing the operation to be as close to
one as possible.  In fact, what one would like, is, given a set of
operations that one wishes to perform, a procedure for systematically
increasing the probability of successfully performing these operations.

In the case of one-parameter unitary groups acting qubits this was done
Preskill \cite{Preskill1998} and Vidal, Masanes
and Cirac (VMC) \cite{Vidal2002}.  Vidal, Masanes and
Cirac considered the one-parameter group of operations given by
$U(\alpha )=\exp (i\alpha\sigma_{z})$, for $0\leq \alpha <2\pi$,
and discussed two equivalent methods of making the probability of
performing $U(\alpha )$ arbitrarily close to one.  A circuit
consisting of a single Controlled-NOT ({\tt CNOT}) gate, with the control qubit
as the data and the target qubit as the program, can successfully
perform $U(\alpha )$ with a probability of $1/2$.  If the procedure
fails, however, the data qubit, which was initially in the state
$|\psi\rangle$, is left in the state $U(-\alpha )|\psi\rangle$.
What we can now do, is to send this qubit back into the same circuit,
but with the program state that encodes the operation $U(2\alpha )$.
This also has a probability of $1/2$ of succeeding, and increases the
total success probability for the two-step procedure to $3/4$.  Note
that our program state has increased to two qubits, one for the first
step and one for the second.  We can continue in this way simultaneously
increasing the success probability and the size of the program state.
It is also possible to design more complicated circuits that perform
the entire procedure at once, i.e.\ they have a one-qubit data state,
an $N$-qubit program state, and a success probability of $1-(1/2)^{N}$
\cite{Vidal2002}.

Here we would like to extend these ideas in a number of different
directions.  First, we shall show that it is possible to boost the
probability of sets of nonunitary operations.  It will then be shown
how to increase the success probability of operations on qudits.
Finally, more complicated groups of operations will be considered.

\section{Operations on qubits}
\label{sec2}
We shall begin by describing the methods developed in \cite{Preskill1998}
and \cite{Vidal2002} in terms of the formalism presented in \cite{Hillery2002b}.
There, the input data state is in the Hilbert space ${\mathcal H}_{d}$,
the program state in the space ${\mathcal H}_{p}$, and $G$ is the unitary
operator, acting on the space ${\mathcal H}_{d}\otimes{\mathcal H}_{p}$,
that describes the action of the circuit.  This operator can be
expressed as
\begin{equation}
\label{decomp}
G=\sum_{j,k=0}^{N}A_{jk}\otimes |j\rangle_{p}\,_{p}\langle k| ,
\end{equation}
where $N$ is the dimension of ${\mathcal H}_{p}$, $A_{jk}$ is an operator
on ${\mathcal H}_{d}$, and $\{ |j\rangle |j=1,\ldots N\}$ is an
orthonormal basis for the program space.  The operators $A_{jk}$ satisfy
\cite{Hillery2002b}
\begin{equation}
\label{unitary}
\sum_{j=1}^{N}A^{\dagger}_{jk_{1}}A_{jk_{2}}=\sum_{j=1}^{N}A_{k_{1}j}
A^{\dagger}_{k_{2}j}=I_{d}\delta_{k_{1}k_{2}} ,
\end{equation}
where $I_{d}$ is the identity operator on ${\mathcal H}_{d}$.  If the
circuit acts on the input state $|\psi\rangle_{d}\otimes |\Xi\rangle_{p}$,
we find that
\begin{equation}
G(|\psi\rangle_{d}\otimes |\Xi\rangle_{p})=\sum_{j=1}^{N}A_{j}(\Xi )
|\psi\rangle_{d}\otimes |j\rangle_{p} ,
\end{equation}
where
\begin{equation}
A_{j}(\Xi )=\sum_{k=1}^{N}\,_{p}\langle k|\Xi\rangle_{p}A_{jk} .
\end{equation}

Let us begin by using this formalism, let us look at a {\tt CNOT}
gate and the simplest of the circuits discussed in \cite{Vidal2002}.
Both the data and program space are two-dimensional, and the
data space is the control qubit and the program space is the
target qubit.  Expressing the operator for the {\tt CNOT} gate in
the form given in Eq.\ (\ref{decomp}), and choosing the basis
$\{ |0\rangle ,|1\rangle \}$ for the program space, we find
that
\begin{eqnarray}
\begin{array}{ll}
A_{00}=|0\rangle\langle 0|; &   A_{01}=|1\rangle\langle 1|; \\
A_{10}=|1\rangle\langle 1|; & A_{11}=|0\rangle\langle 0| .
\end{array}
\end{eqnarray}
We want to use this circuit to perform the operation $U(\alpha )$
and this can be done with the program state
\begin{equation}
|\Xi (\alpha )\rangle = \frac{1}{\sqrt{2}}(e^{i\alpha}|0\rangle
+e^{-i\alpha}|1\rangle ) .
\end{equation}
This gives us the output state
\begin{equation}
\label{out1}
G(|\psi\rangle_{d}\otimes |\Xi (\alpha )\rangle_{p})=\sum_{j=0}^{1}
A_{j}(\alpha )|\psi\rangle_{d}\otimes |j\rangle_{p}
\end{equation}
where the program operators are
\begin{eqnarray}
\label{examp}
A_{0}(\alpha ) & = & \frac{e^{i\alpha}}{\sqrt{2}}|0\rangle
\langle 0|+\frac{e^{-i\alpha}}{\sqrt{2}}|1\rangle\langle 1|
=\frac{1}{\sqrt{2}}U(\alpha )
\nonumber \\
A_{1}(\alpha ) & = & \frac{e^{i\alpha}}{\sqrt{2}}|1\rangle
\langle 1|+\frac{e^{-i\alpha}}{\sqrt{2}}|0\rangle\langle 0|
=\frac{1}{\sqrt{2}}U(-\alpha ).
\end{eqnarray}
Therefore, if we measure the output of the program register in the
computational basis and obtain $|0\rangle$, then $U(\alpha )$ has
been carried out on the data state.  This occurs with a probability
of $1/2$.

If we obtain $|1\rangle$ instead of $|0\rangle$ when we measure the
program register output, then the operation $U(-\alpha )$ has been
performed on the data state.  We can try to correct this by sending
the state $U(-\alpha )|\psi\rangle_{d}$ back into the same circuit,
but with the program state $|\Xi (2\alpha )\rangle_{p}$.  If we measure
the program output and obtain $|0\rangle$, then the output of the
data register is
\begin{equation}
U(2\alpha )U(-\alpha )|\psi\rangle_{d}=U(\alpha )|\psi\rangle_{d} ,
\end{equation}
and this happens with a probability of 1/2.  This will correct the
previous error.

A circuit that does this all at once can be constructed from three
qubits and two quantum gates \cite{Vidal2002}.  Qubit $1$ is the data
qubit, and qubits $2$ and $3$ are the program qubits.  The first
gate is a {\tt CNOT} gate with qubit $1$ as the control and qubit $2$
as the target.  The second gate is a Toffoli gate with qubits $1$
and $2$ as controls and qubit $3$ as the target.  A Toffoli gate
does nothing to the control bits, and does nothing to the target bit
unless both control bits are $1$, in which case it flips the target
bit.  If we denote the orthonormal program space basis by
\begin{eqnarray}
|0\rangle_{p}=|0\rangle_{2}|0\rangle_{3};\qquad & |2\rangle_{p}=|1\rangle_{2}
|0\rangle_{3}\, ; \nonumber \\
|1\rangle_{p}=|0\rangle_{2}|1\rangle_{3};\qquad & |3\rangle_{p}=|1\rangle_{2}
|1\rangle_{3} ,
\end{eqnarray}
then this circuit can be described by the operators
\begin{equation}
\label{2qubit}
\begin{array}{llll}
A_{00}=|0\rangle\langle 0|; & A_{01}=0; & A_{02}=|1\rangle\langle 1|; &
A_{03}= 0; \\
A_{10}= 0; & A_{11}=|0\rangle\langle 0|; & A_{12}=0; &
A_{13}=|1\rangle\langle 1|;  \\
A_{20}=0; & A_{21}=|1\rangle\langle 1|; & A_{22}=|0\rangle
\langle 0|; & A_{23}= 0;  \\
A_{30}= |1\rangle\langle 1|; & A_{31}=0; & A_{32}= 0; & A_{33}=
|0\rangle\langle 0|.
\end{array}
\end{equation}
The program state is now
\begin{equation}
|\Xi (\alpha )\rangle = \frac{1}{2}\sum_{j=0}^{3}e^{i(3-2j)\alpha}
|j\rangle_{p} .
\end{equation}
At the output of the processor
the program register is measured in the computational basis, and
only if both qubits are found to be in the state $|1\rangle$ does the
procedure fail.  The overall probability of succeeding is $3/4$.

Now let us go back to the {\tt CNOT} gate with a single qubit program and
consider a more general program state
\begin{equation}
|\Xi\rangle = c_{0}|0\rangle +c_{1}|1\rangle ,
\end{equation}
the operators $A_{0}(\Xi )$ and $A_{1}(\Xi )$ are
\begin{eqnarray}
A_{0}(\Xi ) & = & c_{0}|0\rangle\langle 0|+c_{1}|1\rangle
\langle 1| \, ;\nonumber \\
A_{1}(\Xi ) & = & c_{1}|0\rangle\langle 0|+c_{0}|1\rangle
\langle 1| .
\end{eqnarray}
These operators are not unitary, but they do have the property that
$A_{0}(\Xi )A_{1}(\Xi )=A_{1}(\Xi )A_{0}(\Xi )=c_{0}c_{1}I$.  The
output state of this circuit is given by Eq.\ (\ref{out1}), so that
it can be used to realize, probabilistically, either of the nonunitary
operators, $A_{0}(\Xi )$ or $A_{1}(\Xi )$.  It also
suggests that we should be able to apply something like the
Preskill-Vidal-Masanes-Cirac scheme.  In
particular, suppose we are trying to perform the operation
\begin{equation}
B(z)=|0\rangle\langle 0|+z|1\rangle\langle 1| .
\end{equation}
If $c_{1}=zc_{0}$, then $A_{0}(\Xi )$ is proportional to $B(z)$.
We send the data state into
the processor and then measure the program state in the $\{
|0\rangle ,|1\rangle\}$ basis.  If we get $0$ we have succeeded,
but if we get $1$ we have instead applied $A_{1}(\Xi )$ to the
state.  If we fail, however, we can try again.  We now take the
output from our first attempt, which is $A_{1}(\Xi )|\psi\rangle_d$,
and send it into the processor again, but this time with the
program state
\begin{equation}
|\Xi^{\prime}\rangle = \left(\frac{1}{1+|z|^{4}}\right)^{1/2}
(|0\rangle + z^{2}|1\rangle ) \, .
\end{equation}
We again measure the program state, and
if we find $0$, the output of the data register is the desired
state, $A_{0}(\Xi )|\psi\rangle_d$.  If we failed, that is we found
$1$, we can try yet again, but we need to modify the program
state every time we repeat the process.

Rather than performing this procedure sequentially, i.e.\
sending in the input state, seeing if we succeed, and if not
trying the procedure again with a modified program state, we
can again do everything at once by {\it enlarging} the size of the program
space.  We shall use a slightly different processor than the one used
by Vidal and Cirac.  It has the same $4$-dimensional program space, but
the operators $A_{jk}$ are now given by
\begin{equation}
\label{2qubit2}
\begin{array}{llll}
A_{00}=|0\rangle\langle 0|; & A_{01}= |1\rangle\langle 1|; & A_{02}=0; &
A_{03}= 0; \\
A_{10}= 0; & A_{11}=|0\rangle\langle 0|; & A_{12}= |1\rangle\langle 1|; &
A_{13}=0;  \\
A_{20}=0; & A_{21}= 0; & A_{22}=|0\rangle\langle 0|; &
A_{23}=  |1\rangle\langle 1|;  \\
A_{30}= |1\rangle\langle 1|; & A_{31}=0; & A_{32}= 0; & A_{33}=
|0\rangle\langle 0|.
\end{array}
\end{equation}
The program state is now
\begin{equation}
|\Xi\rangle_{p}=\sum_{k=0}^{3}c_{k}|k\rangle_{p}\,  ,
\end{equation}
where $c_{k+1}=zc_{k}$ for $k=0,1,2$, and normalization then
requires that
\begin{equation}
|c_{0}|^{2}=\frac{1-|z|^{2}}{1-|z|^{8}} .
\end{equation}
The operation of the processor is given by
\begin{equation}
G(|\psi\rangle_d\otimes |\Xi\rangle_{p})=\sum_{j=0}^{3}
A_{j}(\Xi )|\psi\rangle_d\otimes |j\rangle_{p} ,
\end{equation}
where
\begin{equation}
A_{j}(\Xi )=\sum_{k=0}^{3}c_{k}A_{jk} ,
\end{equation}
and the operators $A_{jk}$ are given in Eq.\ (\ref{2qubit2}).
This processor will perform the operation $B(z)$ with a reasonably
high probability.  In order to see this, we first note that
$A_{j}(\Xi )=z^{j}A_{0}(\Xi )$ for $j=0,1,2,$.  This implies that
\begin{eqnarray}
G(|\psi\rangle_d\otimes |\Xi\rangle_{p}) &=& A_{0}(\Xi )|\psi\rangle_d
\otimes (\sum_{j=0}^{2}z^{j}|j\rangle_{p}) \nonumber \\
& &\ \ \ +
A_{3}(\Xi )|\psi\rangle_d \otimes |3\rangle_{p} ,
\end{eqnarray}
and $A_{0}(\Xi )=c_{0}B(z)$.
At the output of the processor we measure the program state in
the $\{ |j\rangle |j=0,\ldots 3\}$ basis, and if we get $0,1$
or $2$, we have carried out the desired operation.
If $|\psi\rangle_d = \alpha |0\rangle +\beta |1\rangle$, then the
probability of success depends on the input state and is given by
\begin{equation}
P_{suc}=\left(\frac{1-|z|^{6}}{1-|z|^{8}}\right) (|\alpha |^{2}
+|z|^{2}|\beta |^{2}) .
\end{equation}
If we average this probability over all input states we find that
\begin{equation}
\overline{P}_{suc}=\frac{1}{2}\left(\frac{1-|z|^{6}}{1-|z|^{8}}\right)
(1+|z|^{2}) .
\end{equation}
As an example, we can consider the case $|z|^{2}=1/2$, which gives
us $\overline{P}_{suc}=0.7$.

This can easily be generalized to an $N$-dimensional program.  The
operators $A_{jk}$ are now given by
\begin{equation}
A_{jk}=\delta_{j,k}|0\rangle\langle 0|+\delta_{j+1,k}|1\rangle
\langle 1| ,
\end{equation}
where the addition in the second Kronecker delta is done modulo $N$.
These operators satisfy Eq.\ (\ref{unitary}), so that they define a
unitary operator.  The program state is now
\begin{equation}
|\Xi \rangle = c_{0}\sum_{j=0}^{N-1}z^{j}|j\rangle_{p} ,
\end{equation}
where
\begin{equation}
\label{normalize}
|c_{0}|^{2}=\frac{1-|z|^{2}}{1-|z|^{2N}} .
\end{equation}
This yields the following output state
\begin{eqnarray}
G(|\psi\rangle_{d}\otimes |\Xi\rangle_{p}) & = & c_{0}B(z)|\psi\rangle_{d}
\otimes\sum_{j=0}^{N-2}z^{j}|j\rangle_{p} \nonumber \\
 & & \ \ \ +A_{N-1}(\Xi )|\psi_{d}\rangle\otimes |N-1\rangle_{p},
\end{eqnarray}
where
\begin{equation}
A_{N-1}(\Xi ) = c_{0}(z^{N-1}|0\rangle\langle 0|+|1\rangle\langle 1|) .
\end{equation}
The probability of successfully performing $B(z)$ on $|\psi\rangle_{d}$
is given by
\begin{eqnarray}
P_{suc} & = & 1-\| A_{N-1}(\Xi )\psi\|^{2} \nonumber \\
 & = & 1-\frac{(1-|z|^{2})(|\alpha |^{2}|z|^{2(N-1)}+|\beta |^{2})}
{|z|^{2N}-1} .
\end{eqnarray}
When $|z|=1$, this is equal to $1-(1/N)$.  An examination of $P_{suc}$ shows
that it is an increasing function of $N$.  In the case that $|z|=1$ it
approaches $1$ as $N\rightarrow\infty$.  This is no longer true if
$|z|\neq 1$; if $|z|<1$, we find that the limit is
\begin{equation}
P_{suc}\rightarrow 1-(1-|z|^{2})|\beta |^{2}=\| B(z)\psi\|^{2} ,
\end{equation}
and if $|z|>1$, the limit is
\begin{equation}
P_{suc}\rightarrow 1-\left( 1-\frac{1}{|z|^{2}}\right)|\alpha |^{2}
=\frac{1}{|z|^{2}}\| B(z)\psi\|^{2} .
\end{equation}
Therefore, only in the case that we are implementing a unitary operation
can this sequence of processors achieve a success probability arbitrarily
close to $1$.

\section{Qudits}
\label{sec3}
We now want to see how these arguments can be generalized to higher
dimensional systems, and, for the sake of simplicity, let us start by
examining qutrits.  The data space is now
three-dimensional, and let us take for the operators $A_{jk}$
\begin{eqnarray}
A_{00}=|0\rangle\langle 0|;\quad & A_{01}=|1\rangle\langle 1|; \quad& A_{02}=
|2\rangle\langle 2|; \nonumber \\
A_{10}=|2\rangle\langle 2|;\quad  & A_{11}=|0\rangle\langle 0|;\quad & A_{12}=
|1\rangle\langle 1|; \nonumber \\
A_{20}=|1\rangle\langle 1|;\quad & A_{21}=|2\rangle\langle 2|;\quad & A_{22}=
|0\rangle\langle 0| .
\end{eqnarray}
The general program state is
\begin{equation}
|\Xi\rangle = c_{0}|0\rangle +c_{1}|1\rangle +c_{2}|2\rangle ,
\end{equation}
which gives the program operators
\begin{eqnarray}
A_{0}(\Xi )= c_{0}|0\rangle\langle 0| +c_{1}|1\rangle\langle 1|
+c_{2}|2\rangle\langle 2|\, ; \nonumber \\
A_{1}(\Xi )=c_{0}|2\rangle\langle 2| +c_{1}|0\rangle\langle 0|
+c_{2}|1\rangle\langle 1|\, ; \nonumber \\
A_{2}(\Xi )= c_{0}|1\rangle\langle 1| +c_{1}|2\rangle\langle 2|
+c_{2}|0\rangle\langle 0| .
\end{eqnarray}
The output state is
\begin{equation}
|\Psi_{out}\rangle =\sum_{j=0}^{2}A_{j}|\psi\rangle_{d}\otimes
|j\rangle_{p} ,
\end{equation}
so that if we measure in the program space and get $j$, the output
state of the data register is $A_{j}(\Xi )|\psi\rangle_{d}$.

Suppose we are trying to apply the operator $A_{0}(\Xi )$ to the
input data state.  The probability of succeeding is $\langle\psi |
A_{0}^{\dagger}(\Xi )A_{0}(\Xi )|\psi\rangle$.  If we fail, however,
we can try again, and this will increase the total probability of
success.  To see how this works, let us consider an example.  Suppose
that we measured the program register and got $1$ instead of $0$.  That
means we now have the state $A_{1}(\Xi )|\psi\rangle_{d}$.  We take
this state and put it through the processor again, but with a modified
program state
\begin{equation}
|\Xi^{\prime}\rangle = c_{0}^{\prime}|0\rangle +c_{1}^{\prime}|1\rangle
+c_{2}^{\prime}|2\rangle .
\end{equation}
Suppose we now measure the output in the program space and get $0$. If
$A_{0}(\Xi^{\prime})A_{1}(\Xi )\propto A_{0}(\Xi )$, then we have
succeeded on our second try.  Noting that
\begin{equation}
A_{0}(\Xi^{\prime})= c_{0}^{\prime}|0\rangle\langle 0| +c_{1}^{\prime}
|1\rangle\langle 1|  +c_{2}^{\prime}|2\rangle\langle 2| ,
\end{equation}
we see that this condition is satisfied if
\begin{equation}
c_{0}^{\prime}=\frac{\alpha c_{0}}{c_{1}};\quad
c_{1}^{\prime}=\frac{\alpha c_{1}}{c_{2}};\quad
c_{2}^{\prime}=\frac{\alpha c_{2}}{c_{0}} .
\end{equation}
The constant $\alpha$ is chosen so that $|\Xi^{\prime}\rangle$ is
normalized.

What we can conclude from this is that we can, by trial and correction,
boost the probabilities of implementing operators that
are {\it diagonal} in the basis $\{ |0\rangle , |1\rangle , |2\rangle \}$.
In the case that the operator we are trying to implement is unitary,
i.e.\ $|c_{j}|=1/\sqrt{3}$, then our probability of success at each trial
is $1/3$, so that our probability of success after $N$ trials is
$1-(2/3)^{N}$.  This probability goes to $1$ as $N$ goes to infinity.
These conclusions generalize in a straightforward way to qudits.

We now want to explore increasing the probability of successfully
performing an operation on qudits by increasing the size of the program
space.  The data space is now of dimension $D$, and the orthonormal
basis spanning it is $\{ |0\rangle_{d} ,\ldots |D-1\rangle_{d}\}$. We
shall consider a particular kind of operation, one that changes the
amplitude of one of the basis states, and leaves the rest alone (up
to overall normalization).  Suppose the state whose amplitude we
want to change is $|0\rangle_{d}$.  The operator we want to implement is
\begin{equation}
B_{0}(z)=z|0\rangle_{p}\,_{p}\langle 0|+X ,
\end{equation}
where
\begin{equation}
X=\sum_{k=1}^{D-1}|k\rangle_{p}\,_{p}\langle k| .
\end{equation}

For our processor, we shall choose the operators $A_{jk}$, where
$j$ and $k$ run from $0$ to $D-1$ to be
\begin{equation}
A_{jk}=\delta_{jk}X+\delta_{k,j+1}|0\rangle_{p}\,_{p}\langle 0| ,
\end{equation}
were all additions are modulo $D$.  The program state
\begin{equation}
|\Xi\rangle_{p}=c_{0}\sum_{k=0}^{N-1}z^{k}|k\rangle_{p} ,
\end{equation}
where $|c_{0}|^{2}$ is given by Eq.\ (\ref{normalize}), gives us,
for $0\leq j \leq N-2$
\begin{equation}
A_{j}(\Xi )=c_{0}z^{j}B_{0}(z) .
\end{equation}
The probability of successfully performing $B_{0}(z)$ on the data state
$|\psi\rangle_{d}$, $P_{suc}$, is
\begin{equation}
P_{suc}=\frac{|z|^{2(N-1)}-1}{|z|^{2N}-1}\| B_{0}\psi\|^{2} ,
\end{equation}
when $z|\neq 1$, and it is $(N-1)/N$ when $|z|=1$.  In the limit that
$N$ goes to infinity, $P_{suc}$ goes to one if $|z|=1$.  If $|z|>1$
we have that
\begin{equation}
P_{suc}\rightarrow \frac{1}{|z|^{2}}\| B_{0}\psi\|^{2} ,
\end{equation}
and if $|z|<1$, then
\begin{equation}
P_{suc}\rightarrow \| B_{0}\psi\|^{2} .
\end{equation}
As before, we see that it is only in the case that the operation is
unitary that the probability goes to one.

If we want to modify more than one basis vector amplitude, we can apply
these processors successively, each designed to modify a single amplitude.
In the case that all of the operations are unitary, this is a $D$-dimensional,
programmable phase gate, whose probability of succeeding can be made
arbitrarily close to one.

\section{Realization of SU(2) rotations}
\label{sec4}
In the Vidal-Masanes-Cirac model the angle  of the U(1) rotation that is
supposed to be performed on a qubit is encoded in a quantum state of the program.
The rotation itself is then applied on the data qubit via the {\tt CNOT} gate that plays
the role of a programmable processor. As we have discussed  above
the probability of success of the rotation can be
enhanced, providing the data qubit is processed conditionally  in loops.
The dynamics of each ``run'' of the processor is conditioned by the result of the measurement
performed on the program register.

In what follows we will show that an analogous strategy can be applied in the case of
the SU(2) rotations of a qubit, when the parameters (angles) of the SU(2) rotations
are encoded in the state of the program. In our earlier work \cite{Hillery2002b} we have shown
an arbitrary single-qubit
unitary transformation can be implemented with the probability $p=1/4$
by using a
quantum information
distributor machine ({\tt QID}) as the processor. The {\tt QID} is a quantum processor
with a single data qubit and two program qubits. The quantum information distribution is
realized via a sequence of four {\tt CNOT} gates, such that firstly the data qubit controls the
NOT operation on the first and the second program qubits and then the first and the second program
qubits act as the control with the data qubit as the target. At the end of this process a
projective measurement
on the two program qubits is performed. The measurement is performed in the basis:
$\{|0\r|+\r; |0\r|-\r; |1\r|+\r ;|1\r|-\r\}$ (where $|\pm\r = (|0\r\pm|1\r)/\sqrt{2}$). The
realization of the desired transformation is associated with the projection
onto the vector $|0\r|+\r$. In what follows we will explicitly
show how to correct the cases of wrong results, i.e. of projections onto
one of the vectors $|0\r|-\r,|1\r|+\r,|1\r|-\r$.

The action of the {\tt QID} processor is given by relation \cite{Braunstein2001,Hillery2002b}
\be
G=\sum_{j=0}^3\sigma_j\otimes|\Xi_j\r\l\Xi_j|\, ,
\ee
where $\sigma_j$ are standard $\sigma$-matrices with $\sigma_0=\openone$.
The basis program vectors $|\Xi_j\r$ form the standard Bell basis, i.e.
\be
\nonumber
|\Xi_0\r=\frac{1}{\sqrt{2}}(|00\r+|11\r);\quad  &
|\Xi_x\r=\frac{1}{\sqrt{2}}(|01\r+|10\r); \\
\nonumber
|\Xi_z\r=\frac{1}{\sqrt{2}}(|00\r-|11\r);\quad  &
|\Xi_y\r=\frac{1}{\sqrt{2}}(|01\r-|10\r).
\ee
The general program state $|\Xi(\vec{\mu})\r_p$ encoding the
unitary transformation
$U_{\vec{\mu}}=\exp(i\vec{\mu}.\vec{\sigma})=\cos\mu\openone+
i\sin\mu\frac{\vec{\mu}}{\mu}.\vec{\sigma}$ ($\mu=|\vec{\mu}|$)
is given by the expression
\be
|\Xi(\vec{\mu})\r_p=\cos\mu|\Xi_0\r+i\frac{\sin\mu}{\mu}
(\mu_x|\Xi_x\r+\mu_y|\Xi_y\r+\mu_z|\Xi_z\r)\, .
\ee
Performing the previously mentioned measurement in the program basis
$|0+\r,|0-\r,|1+\r,|1-\r$ we obtain  the following unitary transformations
\be
\nonumber
|0\r\otimes|+\r & : & |\psi\r_d\to U_{\vec{\mu}}|\psi\r_d \, ;\\
\nonumber
|0\r\otimes|-\r & : & |\psi\r_d\to \sigma_z U_{\vec{\mu}}\sigma_z|\psi\r_d \, ;\\
\nonumber
|1\r\otimes|+\r & : & |\psi\r_d\to\sigma_x U_{\vec{\mu}}\sigma_x|\psi\r_d \, ; \\
\nonumber
|1\r\otimes|-\r & : & |\psi\r_d\to\sigma_y U_{\vec{\mu}}\sigma_y|\psi\r_d\, ,
\ee
where
\be
\label{qidd}
U_{\vec{\mu}}=
\cos\mu\openone+\frac{i\sin\mu}{\mu}(\mu_x\sigma_x+\mu_y\sigma_y+\mu_z\sigma_z)\, .
\ee
To obtain this simple expression we have used the identity
$\sigma_j\sigma_k\sigma_j=-\sigma_k$ if $k\ne j$.
All observed outcomes  occur with the same probability, $p=1/4$.
Using the above notation the action of the {\tt QID} can be expressed
in the form
\be
|\psi\r_d\otimes|\Xi(\vec{\mu})\r_p\to\frac{1}{2}\left(
\sum_{j=0}^3\sigma_j U_{\vec{\mu}}\sigma_j|\psi\r_d\otimes|\tilde{j}\r_p\right)
\ee
where vectors $\{|\tilde{j}\r_p\}$ form the basis of $\cH_p$
associated with the realized measurement. The explicit form of the vectors
is presented in following Section where we discuss a general solution of
SU(N) rotations of qudits.

We see that each outcome of the measurement indicates a different unitary
transformation has been applied to the data.
Once we have obtained a specific result we can use the same processor
again to correct an incorrectly transformed data register and consequently
improve the success probability. In particular, in the case
of the result $j$, the new program register
needs to encode the correcting transformation
$U^{(1)}_j=U_{\vec{\mu}}\sigma_j U_{\vec{\mu}}^\dagger\sigma_j$.
The probability of implementing the unitary transformation using
one conditioned loop is given as
$p(1)=\frac{1}{4}+3\frac{1}{16}=\frac{7}{16}$. Using more and more
conditioned loops the success probability is given by
$p(n)=\sum_{j=1}^n \frac{1}{4^j}3^{j-1}=\frac{1}{4}\sum_j
(\frac{3}{4})^j=\frac{1}{4}\frac{1-(3/4)^n}{1/4}=1-(3/4)^n$
converges to unity, i.e. $p(n)\to 1$ as the number of conditioned loops
$n$ goes to infinity. For instance, thirty conditioned loops result
in the negligible probability of failure, $p\simeq 10^{-4}$.

The example of Vidal, Masanes and Cirac  shows us that we are able to replace the
feedback scenario with a probabilistic scenario by using different
processors. An open problem is whether the same
replacement can be done in general, or at least for the case
of the {\tt QID}.

\section{SU (N) rotations of qudits}
\label{sec5}

In what follow we will show that one can utilize the {\tt QID} for a probabilistic
implementation of SU(N) rotations of qudits. We start our discussion with a brief description of
the {\tt QID} in the case of qudits. First, we introduce a generalization of the
two-qubit {\tt CNOT} gate \cite{Braunstein2001} for qudits. This is a conditional shift operator
defined with a control qudit ``{\em a}'' and the target qudit ``{\em b}''
\begin{eqnarray}
\label{3.3}
{D}_{ab}=
\sum_{k,m=0}^{N-1} |k\rangle_a\langle k|\otimes
|(m+k){\rm mod}\,N\rangle_b\langle m|\; ,
\end{eqnarray}
which implies that
\begin{eqnarray}
{D}_{ab}^{\dagger}=
\sum_{k,m=0}^{N-1} |k\rangle_a\langle k|\otimes
|(m-k){\rm mod}\,N\rangle_b\langle m|\; .
\end{eqnarray}
>From this definition it follows that the operator
${D}_{ab}$ acts on the basis vectors of a qudit as
\be
{D}_{ab}|k\r|m\r = |k\r|(k+m) {\rm mod}\, N\r\;,
\label{3.4}
\ee
which means that this operator has the same action as
the conditional adder and can be performed with the help of
the simple quantum network discussed in \cite{Vedral1996}.
Note  that for $N>2$
the two operators $ D$ and ${D}^\dagger$  differ;
they describe conditional shifts in opposite directions.
Therefore the generalizations of the {\tt CNOT} operator to higher
dimensions are just {\em conditional shifts}.

Following our earlier work \cite{Hillery2002b,Braunstein2001} we can
assume the network for the probabilistic universal quantum
processor to be
\be
{P}_{123}=
{D}_{31}{D}_{21}^\dagger{D}_{13}{D}_{12}\; .
\label{3.5}
\ee
The data register consists of system $1$ and the program register of
systems $2$ and $3$.  The state $|\Xi_V\rangle_{23}$ acts as
the ``software'' that caries the information about
the operation $V$  to be implemented on the
qudit data state $|\Psi\r_1$.  The output state
of the three qudit system, after the four controlled
shifts are applied, reads
\begin{eqnarray}
\label{3.6}
|\Omega\rangle_{123}=
{D}_{31}{D}_{21}^\dagger{D}_{13}{D}_{12}
|\Psi\rangle_1|\Xi_V\rangle_{23}\;.
\end{eqnarray}

The sequence of four operators acting on the basis vectors
gives
$|n\rangle_{1}|m\rangle_{2}|k\rangle_{3}$ as
\end{multicols}
\vspace{-0.2cm}
\noindent\rule{0.5\textwidth}{0.4pt}\rule{0.4pt}{0.6\baselineskip}
\vspace{0.2cm}
\begin{eqnarray}
\label{3.7}
{D}_{31}{D}_{21}^\dagger{D}_{13}{D}_{12}
|n\rangle_{1}|m\rangle_{2}|k\rangle_{3}=
|(n-m+k){\rm mod}\,N\rangle_{1}\,|(m+n){\rm mod}\,N\rangle_{2}
\,|(k+n){\rm mod}\,N\rangle_{3}\;.
\end{eqnarray}
  \hfill\noindent\rule[-0.6\baselineskip]%
  {0.4pt}{0.6\baselineskip}\rule{0.5\textwidth}{0.4pt}
\vspace{-0.2cm}
\begin{multicols}{2}
We now turn to the fundamental program states.
A basis consisting of maximally entangled two-particle states
(the analogue of the Bell basis for spin-$\frac{1}{2}$ particles)
is given by
\begin{eqnarray}
\label{3.8}
|\Xi_{mn}\rangle = \frac{1}{\sqrt{N}} \sum_{k=0}^{N-1} \exp \Bigl(
i\frac{2\pi}{N} mk \Bigr) |k\rangle|(k-n){\rm mod}\,N\rangle \,,\!\!\!
\end{eqnarray}
where $m,n=0,\dots,N-1$.  If $|\Xi_{mn}\rangle_p$ is the initial
state of the program register,
and $|\Psi\rangle=\sum_j \alpha_j |j\rangle_d$
(here, as usual,  $\sum_j |\alpha_j|^2=1$) is the initial state of
the data register, then follows that
\be
&& {P}_{123}|\Psi\rangle_{1}|\Xi_{mn}\rangle_{23} =
\nonumber\\
&=&
\sum_{jk}\frac{\alpha_j}{\sqrt{N}}\exp\left[\frac{2\pi ikm}{N}\right]
{P}_{123}
|j\rangle |k\rangle |k-n\rangle
\nonumber\\
&=&\sum_{jk}\frac{\alpha_j}{\sqrt{N}}\exp\frac{2\pi ikm}{N}
|j-n\rangle |k+j\rangle |k+j-n\rangle
\nonumber\\
&=&\sum_{jk}\alpha_j\exp\frac{-2\pi ijm}{N}|j-n\rangle|\Xi_{mn}\rangle
\nonumber\\
&=&(U^{(m,n)}|\Psi\rangle)|\Xi_{mn}\rangle ,
\label{3.9}
\ee
where we have introduced the notation
\be
\label{3.10}
U^{(m,n)}=\sum_{s=0}^{N-1}\exp\frac{-2i \pi sm}{N}
|s-n\rangle\langle s|.
\ee
This result is similar to the one we found in the case of a
single qubit (see previous section).
The operators $U^{(m,n)}$ satisfy the orthogonality relation
\begin{equation}
\label{orth}
{\rm Tr}\left[ (U^{(m^{\prime},n^{\prime})})^{\dagger}U^{(m,n)}
\right] = N\delta_{m,m^{\prime}}\delta_{n,n^{\prime}} .
\end{equation}
 The space of linear operators ${\cal T}({\cal H})$
defined on some Hilbert space $\cal H$
with the scalar product given by (\ref{orth}) we know as
{\it Hilbert-Schmidt space}. Thus the unitary operators $U^{(m,n)}$ form
an orthogonal basis in it
and any operator $V\in{\cal T}({\cal H})$ can be expressed in terms of them
\begin{equation}
\label{expans}
V=\sum_{m,n=0}^{N-1}d_{mn}U^{(m,n)} .
\end{equation}

The orthogonality relation allows us to find the expansion coefficients
in terms of the operators
\be
d_{mn}=\frac{1}{N}{\rm Tr}\left[\left(U^{(m,n)}\right)^\dagger V
\right].
\ee
Therefore, the program vector that implements the operator $V$ is
given by
\begin{equation}
|\Xi_{V}\rangle_{23} =
\sum_{m,n=0}^{N-1}d_{mn}|\Xi_{mn}\rangle_{23} .
\end{equation}
Application of the processor to the input state $|\Psi\rangle_{1}
|\Xi_{V}\rangle_{23}$ yields the output state
\be
|\Omega\rangle_{123}=
\sum_{mn}d_{mn}U^{(m,n)}|\Psi\rangle_{1}\otimes|\Xi_{mn}\rangle_{23}.
\label{3.22}
\ee

Now let us perform a measurement of the program output in the basis
\be
|\Phi_{rs}\r =\frac{1}{N} \sum_{m,n=0}^{N-1}
\exp\left[2\pi i\frac{(m r - n s)}{N}\right]
|\Xi_{mn}\r \, .
\label{3.23}
\ee
The orthogonality of this measurement basis directly follows from the orthogonality of the entangled basis
$|\Xi_{mn}\r$. We should also note, that the vectors $|\Phi_{rs}\r$ itself can be rewritten in a factorized
form, i.e.
\be
|\Phi_{rs}\r = |-r\r \otimes \frac{1}{\sqrt{N}}\sum_{n=0}^N \exp\left[2\pi i\frac{n s}{N}\right]
|n-r\r\, ,
\label{3.24}
\ee
which means that the measurement can be performed independently on two program qudits.

In order to clarify the role of the measurement we will rewrite the output state of the {\tt QID} using
the basis $|\Phi_{rs}\r$ for program qudits:
\end{multicols}
\vspace{-0.2cm}
\noindent\rule{0.5\textwidth}{0.4pt}\rule{0.4pt}{0.6\baselineskip}
\vspace{0.2cm}
\be
{P}_{123}|\Psi\rangle_{1}|\Xi_{V}\rangle_{23} & =&
\sum_{m,n=0}^{N-1} d_{m,n} U^{(m,n)}
|\Psi\rangle_1 |\Xi_{mn}\rangle_{23}
\nonumber\\
&=&
\sum_{m,n=0}^{N-1} d_{m,n} U^{(m,n)}
|\Psi\rangle_1 \left[\frac{1}{N} \sum_{r,s=0}^{N-1}
\exp\left[-2\pi i\frac{(m r - n s)}{N}\right]
|\Phi_{rs}\r_{23}\right]
\nonumber\\
&=&
\frac{1}{N}\sum_{r,s=0}^{N-1}
\sum_{m,n=0}^{N-1}
\left\{\exp\left[-2\pi i\frac{(m r - n s)}{N}\right]
d_{m,n} U^{(m,n)}\right\}
|\Psi\r_{1}
|\Phi_{rs}\rangle_{23}   .
\label{3.25}
\ee
Taking into account that
\be
\left[U^{(p,q)}\right]^\dagger U^{(m,n)} U^{(p,q)} =
\exp\left[2\pi i\frac{(m q - n p)}{N}\right] U^{(m,n)}
\label{3.26}
\ee
and choosing $p=s$ and $q=r$ we find
\be
\frac{1}{N} {\rm Tr}\left[ \left(U^{(s,r)}\right)^\dagger \left(U^{(m,n)}\right)^\dagger
U^{(s,r)} V\right]
= \exp\left[-2\pi i\frac{(m r - n s)}{N}\right] \, d_{m,n} \, .
\label{3.27}
\ee
Finally, the output of the {\tt QID} can be rewritten in the form
\be
 {P}_{123}|\Psi\rangle_{1}|\Xi_{V}\rangle_{23} =
\frac{1}{N}\sum_{r,s=0}^{N-1}
\left[ U^{(s,r)} V \left(U^{(s,r)}\right)^\dagger \right]
|\Psi\rangle_1 |\Phi_{rs}\r_{23}\, ,
\label{3.28}
\ee
\hfill\noindent\rule[-0.6\baselineskip]%
  {0.4pt}{0.6\baselineskip}\rule{0.5\textwidth}{0.4pt}
\vspace{-0.2cm}
\begin{multicols}{2}
\noindent
from which it is clear that if the result of the measurement of the two program qudits is
$|\Phi_{rs}\r_{23}$, then  the system (data) is left in the state
$\left[ U^{(s,r)} V \left(U^{(s,r)}\right)^\dagger \right]
|\Psi\rangle_1 $. Obviously, if $s=r=0$, then the operator $V$ is applied on the data qudit.
The probability of this outcome is $1/N^2$. For all other results of the measurement
the data qudit is left in the state given above. One can use these output states with a modified
program state to improve the performance of the programmable processor. Specifically, we have to use
the new program state $|\Xi_V^{(r,s)}\r$ that is chosen
after taking into account
the result of the previous
measurement. This program state has first to ``correct'' the wrong realization of the
operation $V$ during the previous ``run'' of the processor
and then apply (probabilistically), the original operation $V$.
For this reason, the new program state has to perform the operation
\be
V^{(r,s)} =
 V \left[
 U^{(s,r)} V \left(U^{(s,r)}\right)^{\dagger}\right]^{-1}\, .
\label{3.29}
\ee
This process of error correction (conditional loops) can be used $K$ times and
the technique of conditioned loops
can be exploited in order to amplify the probability of success.
Applying the processor $K$ times the probability of a successful application
of the desired SU(N) operation $V$
reads $p(K)=1-(1-1/N^2)^K$.

\section{Conclusions}
In this paper we have analyzed a probabilistic programmable quantum processor. We have shown
how to encode information about the
quantum dynamics $V$ to be performed on  a quantum system (data register)
in the state of another quantum system (program register). This information is
stored  in such a way that the program can be used to probabilistically perform
the stored transformation on the data. In our paper we have analyzed
systematically how to perform U(1) rotations of qubits and qudits
and one-parameter families of nonunitary operations
when the angle of rotation
is encoded in states of quantum programs.
In addition we have shown  how to increase
the probability of success when the quantum processor is used in loops with updated program
states. We have generalized the whole problem and we have shown that one can use a very simple
quantum processor, the so called quantum information distributor, to perform arbitrary SU(2) rotations
of qubits as well as SU(N) rotations of qubits using the probabilistic programmable processor
with the quantum program register initially prepared in states that carry the information about
the operation to be performed on the data.
It is also possible to use enlarged programs to increase the probability of success without the use
of loops. In this case the measurement performed on the program register has to be modified accordingly.
We have shown that if the processor
is used in loops with properly chosen program states one can improve the performance of the quantum
programmable processor so that the probability of failure decreases exponentially with the
number of program qudits that store the information about transformation on the data qudit.

\acknowledgements
This was work supported in part
by  the European Union projects QGATES  and
CONQUEST,
by the  National Science Foundation under grant
PHY-0139692, and by  the Slovak Academy of Sciences.


\end{multicols}
\end{document}